\documentclass[12pt]{article}
\usepackage{epsfig}

\newcommand{\be}{\begin{equation}}
\newcommand{\ee}{\end{equation}}

\newcommand{\gamde}{$\gamma d\to\pi^0 d$}
\newcommand{\bfpide}{\mbox {\boldmath $\pi^- d\to\pi^- d$}}

\newcommand{\bfp}{{\bf p}}
\newcommand{\bfk}{{\bf k}}
\newcommand{\bfs}{{\bf s}}
\newcommand{\bfq}{{\bf q}}
\newcommand{\bfr}{{\bf r}}

\newcommand{\bftau}{\mbox {\boldmath $\tau$}}

\newcommand{\bfi}{{\bf I}}

\newcommand{\mkp}{\mbox {$m {\bf k} -\mu {\bf p}$}}
\newcommand{\mmu}{\mbox {$m +\mu$}}

\date{}

\begin{document}


\centerline{\Large \bf On rescattering effects in the reaction \bfpide}

\vspace{1.5cm}

\centerline{\large V.V.~Baru$^{a,b}$, A.E.~Kudryavtsev$^b$, V.E.~Tarasov$^b$}

\vspace{0.3cm}

\begin{center}
\it$^a$Institut f\"{u}r Kernphysik, Forschungszentrum J\"{u}lich
GmbH,\\ D--52425 J\"{u}lich, Germany\\
$^b$Institute of Theoretical and Experimental Physics, \\
117259, B. Cheremushkinskaya 25, Moscow, Russia 
\end{center}










\begin{abstract}

We discuss rescattering corrections to the impulse approximation for the
processes $\pi^- d\to\pi^- d$ and  $\gamma d\to\pi^0 d$.
It is shown that the rescattering effects (RE) give non-negligible
contribution to the real part of these amplitudes. 
At the same time the contributions from the imaginary parts of 
impulse and rescattering corrections drastically cancel each other. 
This cancellation means that the processes $\pi^- d\to \pi^0 nn$ and 
$\gamma d\to \pi^+ nn / \pi^- pp$ are strongly suppressed near
threshold as required by the Pauli principle.


\end{abstract}


\section{Introduction}

The study of the reactions $\gamma d\to \pi^0 d$ and $\pi^- d\to \pi^- d$ near threshold 
has attracted continuous attention in the past few decades. Moreover
the new experimental data appeared due to recent success of the accelerator technologies 
stimulate increasing theoretical interest in this field. In this paper we would like to concentrate
on the rescattering effects (RE) and their role for these reactions. 
Indeed, these effects 
are found to be important in many of theoretical investigations of the reaction  $\gamma d\to \pi^0 d$ (cf.,
e.g. Refs. \cite{2,3,4,Beane}). However, recently in Ref. \cite{1} the discussion about the role of these 
effects was renewed.
In particular  it was emphasized in~\cite{1} that the contribution from the
two-step process  $\gamma d\to \pi^- pp\to \pi^0 d$ (cf. Fig. \ref{1}a) is totally compensated
by the loop corrections to the impulse approximation (LCIA) (cf. Fig. \ref{1}b).
This was argued by the  Pauli principle for the intermediate NN states.
Thus, the rescattering effects  in Ref. \cite{1} do not contribute
to the process of coherent $\pi^0$ photoproduction on deuteron near
threshold. Obviously, this conclusion of Ref.~\cite{1} disagrees with the 
results of other calculations performed, e.g., in  Refs.~\cite{2,3,4,Beane}.

Let us discuss the arguments of Ref. \cite{1} in more detail:

\begin{itemize}
\item[i)]
The final $\pi^0 d$ state has quantum numbers $J^P = 1^-$ at low
energies where pion is in the $S$ wave with respect to the deuteron.
However, the only possible 
state of the system $pp\pi^-$ with
$l_1 = l_2 =0$ is $0^-$ (here $l_1$ is the orbital angular momentum
of the $pp$ system and $l_2$ is the orbital momentum of the pion
relative to the $pp$ system). Therefore  the S-wave
intermediate  state $pp\pi^-$  does not contribute to the process \gamde.

\item[ii)]
In other words, the contribution of the diagram in Fig. \ref{1}a
has to be compensated by the loop
corrections to the impulse approximation (Fig. \ref{1}b) because of 
antisymmetry of the wave function for the pair of the intermediate nucleons.
\end{itemize}


Note, that the process $\gamma d\to\pi^0 np\to \pi^0 d$ is allowed by quantum numbers. 
However the amplitude $\gamma n\to \pi^0 n$ which contributes to this reaction is about factor of 20
smaller than the corresponding amplitude for the charged pion production.

In this remark we are going to discuss the  role of
rescattering effects for the process of pion-deuteron
elastic scattering at low energies. 
The diagrams corresponding to RE and LCIA for the $\pi d$-scattering are
very similar to the ones for the reaction $\gamma d\to \pi d$ (cf. Fig. 
\ref{1} and Fig \ref{2}b and \ref{2}c). Therefore we will  investigate 
the relevance of RE and the problem of  the cancellation of RE and LCIA
performing the calculation of the $\pi d$-scattering amplitude. 

The $\pi d$-scattering
length was measured with a high accuracy~\cite{6,7} and its value
coincides with the theoretical predictions (cf., e.g.,
Refs.~\cite{8,9,10,11}). In all these theoretical calculations 
rescattering effects (including the two-step charge exchange process
$\pi^- p\to \pi^0 n\to \pi^- p$) give significant contribution to the
value of the pion-deuteron scattering length.

In what follows we will directly demonstrate that the real part of the rescattering
diagram (cf. Fig.\ref{2}c) gives non-negligible contribution to the 
pion-deuteron scattering length. 
It is not compensated by the real part of LCIA (cf. Fig.\ref{2}b). However, the 
imaginary parts of RE and LCIA cancel each other. This cancellation means that
there is no contribution to observables from the $\pi NN$ states forbidden
by the Pauli principle.

\section{Calculation of the $\pi d$ scattering amplitude}

Below we use a simple 
potential approach for the calculation 
of the $\pi N$-scattering amplitude. 
This approach was already applied to the problem of
the determination of the $\pi N$-scattering length in Ref. \cite{9}. 
The model utilizes a pion-nucleon potential $V_{\pi N}(\bfp, \bfq )$
which is required for solving the Lippman-Schwinger equation

\be
T = V + V G T\, .
\label{1}
\ee

The S-wave $\pi N$-lengths $b_0$ and $b_1$ are related to the scattering length 
$a_{\pi N}$ by the equation

\be
a_{\pi N} = b_0 + b_1\,\bfi \bftau\, ,
\label{2}
\ee
where \bfi\ and \bftau\ are isospin operators for pion and nucleon,
$b_0$ and $b_1$ are isoscalar and isovector scattering lengths.
The analyses~\cite{9,10} of the experimental data~\cite{6,7} show that
the absolute values of $b_0$ and $b_1$ are small compared to the
typical scale of the problem $\sim \mu^{-1}$ (where $\mu$ is the pion  mass). 
Note also that $b_0\ll b_1$. 
Thus, the amplitude $T$ in eq. (\ref{1}) may be perturbatively expanded  in terms of the potential
$V_{\pi N}(\bfp, \bfq )$. 

Following Ref.~\cite{9} we choose $V_{\pi N}$ in the S-wave in the separable form:

\be
V_{\pi N}(\bfk, \bfq ) =
 -\,\frac{(\lambda_0 +\lambda_1\bfi\bftau )}{2m_{\pi N}}\,
g(k) g(q)\, ,
\label{3}
\ee
where $g(k) = (c^2 + k^2)^{-1}$, $m_{\pi N}=m\mu/(m+\mu)$ and $m$ is the
nucleon mass. The cut off parameter $c$ characterizes the range of 
the $\pi N$-forces, and usually it is varied in the range $2.5\mu\le c\le 5\mu$ \cite{9,10}. 
The parameters $\lambda_0$ and $\lambda_1$ are chosen in such a way to reproduce the 
scattering lengths $b_0$ and $b_1\,$. In what follows we will calculate
the pion-deuteron scattering amplitude up to the second order in terms
of the potential $V_{\pi N}$. With this accuracy  $\lambda_0$ and
$\lambda_1$ are equal to

\begin{eqnarray}
\nonumber
\lambda_0 & = &\frac{c^4}{2\pi^2}\,
\left(b_0 -\,\frac{c}{2}\, (b^2_0 + 2 b^2_1)\right)\, , \\
\lambda_1 & = &\frac{c^4}{2\pi^2}\,b_1
\left(1 -\,\frac{c}{2}\, (2 b_0 - b_1)\right)\, .
\label{4}
\end{eqnarray}

Corrections to these expressions are of the order of $\sim O(b_0^3,b_1^3)$
which are negligible.

Let us calculate the pion-deuteron scattering length using the
potential $V_{\pi N}$~ (cf. eq. (\ref{3})).

\vspace{3mm}

i) \underline{Single scattering amplitude in the Born approximation}

\vspace{5mm}

The diagram corresponding to this amplitude is shown in Fig. \ref{2}a. 
The expression for the $\pi d$ amplitude $f^{1(V)}_{\pi d}$ 
corresponding to the sum of two diagrams with the scattering of pion on proton and neutron has the form:

\be
f^{(1)V}_{\pi d} =-\,\frac{\mu}{(2\pi)(1+\mu/m_d)}\,\int d\bfp\,
 \varphi^2_d (\bfp)\, [V_{\pi^- p} +V_{\pi^- n}]\, .
 \label{5}
\ee

Here $\varphi_d (\bfp)$ is the deuteron wave-function in the momentum
space with the normalization condition
$\int d\bfp\, \varphi^2_d (\bfp)\, = (2\pi)^3$.
Neglecting by the small corrections of order of $\sim \mu/m$, one may take out the
potential $V$ in Eq.~(\ref{5}) of the integral and then get:
\be
f^{1(V)}_{\pi d} = 2
\left[\, b_0 -\frac{c}{2}\, (b^2_0 +2b^2_1)\, \right]\, .
\label{6}
\ee

This contribution is real as it should be in the Born approximation.
Note also that the value $f^{1(V)}_{\pi d}$ depends on the value of the
parameter $c$.

\vspace{3mm}

ii) \underline{Single scattering in the one-loop approximation}

\vspace{5mm}

The diagram for the one-loop correction to the Born approximation is shown
in Fig. \ref{2}b. We have to calculate the sum of two diagrams with the
scattering of pion on proton and neutron taking into account the sum
over all intermediate states. The expression for the amplitude 
$f^{(1)VGV}_{\pi d}$ corresponding to this sum has the form:

\begin{eqnarray}
\nonumber
f^{(1)VGV}_{\pi d} &=& \frac{2\mu}{1+\mu/m_d}\left [(\lambda_0^2+\lambda_1^2)I(\Delta m\!=\!0)+\lambda_1^2
I(\Delta m)\right],\\
I(\Delta m)  = &&\hspace*{-1cm} \int \frac{d\bfp}{2\pi}\,
\frac{ g^2\! \left( \frac{\mkp}{\mmu}\! \right)\,
\varphi^2_d (\bfp)}{(2m_{\pi N})^2} 
\int \frac{d\bfs\,\,  g^2
 \left(\frac{\mkp}{\mmu} +\displaystyle\frac{m-\mu}{m+\mu}\bfs \right)}
 {\rule{0pt}{13pt} \left[\,
\displaystyle \frac{(\bfk\! +\!\bfs)^2}{2\mu}\! +\! \frac{p^2}{2m}\! +\! \frac{(\bfp\! -\!\bfs)^2}{2m}
 +\varepsilon_d\! -\!\Delta m\! -\!\frac{k^2}{2\mu}\! -i0\!\, \right]}\, .
 \label{7}
\end{eqnarray}

Here $\bf k$ is the 3-momentum of the initial and final pion,
$\Delta m =m_{\pi^-} +m_p -m_{\pi^0} -m_n = 3.3$~MeV is
the excess energy for the charge exchange process $\pi^- p\to \pi^0 n$ in the
intermediate state. For the case of the elastic rescattering $\Delta m =0$.
%


%
The integral in eq. (\ref{7}) is calculated numerically for some 
values of the cut off parameter $c$. In the limit of large $c$, i.e. when
$c\gg \mu$ and for $\mu/m\ll 1$ this integral can be calculated 
analytically:

\be
f^{(1)VGV}_{\pi d} = c (b^2_0 +2b^2_1) +2i\left [k_0(\Delta m\!=\!0)(b^2_0 +b^2_1)+k_0(\Delta m)b^2_1\right ],
\label{9}
\ee

where we introduced the notation
$k^2_0 =k^2 +2\mu\Delta m -2\mu\varepsilon_d$. Note that $k\ll c$
near the threshold.

Thus, in the limit of large $c$ the resulting contribution 
from the impulse approximation (cf. Fig. \ref{2}a and \ref{2}b)
to the real part of the $\pi d$-scattering amplitude is

\be
{\rm Re}f^1_{\pi d} = f^{1(V)}_{\pi d} + {\rm Re}f^{(1)VGV}_{\pi d}
 = 2\,b_0\, .
\label{10}
\ee

This is a naive but expected result for the real part of
the amplitude corresponding to the impulse approximation. 
The values of $Re f^{(1)VGV}_{\pi d}$ for the charge exchange process $\pi^- d\to \pi^0 nn\to \pi^- d$ 
are presented in Table 1 for the different values of parameter $c$.
In contrary to the real part of the loop amplitude 
the imaginary part of $f^{(1)VGV}_{\pi d}$ (cf. eq. (\ref{9}))
does not depend on  $c$ as required by the unitarity.

Now let us discuss the contribution to the pion-deuteron scattering
length from the double scattering process.

\vspace{3mm}

iii) \underline{Double scattering contribution}

\vspace{5mm}

Double scattering diagram is shown in Fig. \ref{2}c. Performing the calculation 
we have the following integral for the
doublescattering amplitude $f^{(2)}_{\pi d}$ (cf. Ref. \cite{9} for details):

\begin{eqnarray}
\nonumber
f^{(2)}_{\pi d} &=& \frac{4c^4}{(2\pi)^5}\,\left [ (b^2_0 - b^2_1)\; J(\Delta m\!=\!0)- b^2_1\; J(\Delta m)\right ],\\
J(\Delta m)&=&\int \frac{d\bfq_1 d\bfq_2\, \varphi_d (\bfq_1)\, \varphi_d (\bfq_2)\,
g^2(\bfk + \bfq_1 -\bfq_2 )}{(\bfk\! +\! \bfq_1\! -\!\bfq_2)^2 +
(\mu/m)\,(q^2_1 + q^2_2) +
2\mu\, (\varepsilon_d -\! \Delta m ) - k^2 -i0 }\, .
\label{11}
\end{eqnarray}

In the limit of large $c$ and for $\mu/m \ll 1$ this
integral is reduced to the following expression:

\be
f^{(2)}_{\pi d} = 2\,(b^2_0 - b^2_1) \int \Psi^2_d (r)\,\frac{e^{-i\bfk\bfr +ik_0(\Delta m\!=\!0) r}}{r}\,d\bfr\,
- 2b^2_1 \int \Psi^2_d (r)\,\frac{e^{-i\bfk\bfr +ik_0(\Delta m) r}}{r}\,d\bfr\ ,
\label{12}
\ee

where $\Psi_d (\bfr)$ is the deuteron wave function in the coordinate
space.

In the limit of small $k$ and $k_0$, i.e. near the threshold for the real part
of $f^{(2)}_{\pi d}$ we get

\be
Re f^{(2)}_{\pi d} = 2\,(b^2_0 - 2b^2_1)
\left< \frac{1}{r}\right>_d\, .
\label{13}
\ee

This expression is well known as a static limit for the doublescattering
amplitude, see, e.g.~\cite{12} and references therein.

The imaginary part of the amplitude
$f^{(2)}_{\pi d}$~(\ref{12}) in the same limit  is

\be
{\rm Im}f^{(2)}_{\pi d} = 2k_0(\Delta m\!=\!0)\,(b^2_0 - b^2_1)\,-2k_0(\Delta m)\,b^2_1 .
\label{14}
\ee

Note that this contribution is negative because $b_1\gg b_0$.

\vspace{1.5cm}

iv) \underline{Total pion-deuteron amplitude}

\vspace{0.5cm}

Let us discuss the value for total pion-deuteron scattering amplitude
in the limit of large $c$ ($c\gg \mu$) and for $\mu/m \ll 1$. For the
imaginary part of the resulting amplitude in this limit from eqs. (\ref{9})
and (\ref{14}) we get

\be
{\rm Im}f_{\pi d}\approx 
4 k_0(\Delta m\!=\!0) b^2_0\, .
\label{15}
\ee

Thus, we obtain that the contributions from LCIA and RE to the 
imaginary part of the pion-deuteron scattering amplitude cancel each other
in the leading order (i.e. terms $\sim b^2_1$). 
The non-vanishing part of Im$\,f_{\pi d}$ is proportional to
$b^2_0$ what corresponds to the elastic rescattering process $\pi^- d\to \pi^- pn\to \pi^-d$.
Note, that the imaginary parts of both expressions (\ref{9}) and (\ref{14})
behave as two-particle phase space, i.e.
proportional to $k_0\sim Q^{1/2}$, where $Q$ is the  kinetic energy
of the intermediate $\pi NN$ system. However,
three-particle $\pi NN$ phase space should behave as $Q^2$. This paradox can be resolved if we
remind the reader  that the approximation
$\mu/m \ll 1$, which implies that the kinetic
energies of the intermediate nucleons in Eqs. (\ref{7}) and (\ref{11}) are neglected,
was  used to obtain Eqs. (\ref{9}) and (\ref{14}). This approximation
corresponds to the the rescattering of pion on the fixed centers. That is why the
imaginary parts in Eqs. (\ref{9}) and (\ref{14}) behave as $Q^{1/2}$. Of course, one
can avoid this unnecessary simplification and calculate  Eqs. (\ref{7}) and (\ref{11})
 with  taking into account the terms of order of $O(\frac{\mu}{m})$.
But it should not change the main result that the contributions from LCIA and RE to the
imaginary part of the $\pi d$-scattering amplitude with intermediate charge exchange
cancel each other.
The result (\ref{15}) means that the only possible final state which can be formed in the S-wave 
in the process of deuteron desintegration  is $pn\pi^-$ (with $S=1$ and $I=0$ for pair of nucleons).
The virtual charge exchange does not contribute to the imaginary part of
the pion-deuteron amplitude. This conclusion is in agreement with
the remark of Ref.~\cite{1}.

At the same time we would like to stress that
there is no complete cancellation between real parts of
the amplitudes $f^{(1)VGV}_{\pi d}$ and $f^{(2)}_{\pi d}$, i.e. the resulting contribution
from LCIA and RE to the real part of the pion-deuteron scattering amplitude is not small. 
This conclusion, which is also correct for the process $\gamma d\to \pi^0 d$, is in 
contrary to the arguments of Ref. \cite{1}.
As can be seen from Eq.~(\ref{9}), the expression for Re$\,f^{(1)VGV}_{\pi d}$ depends linearly on the cut-off
parameter $c$ for large values of $c$ and $\mu/m\ll 1$, 
whereas  Re$\,f^{(2)}_{\pi d}$ in the same
limit is totally determined by the deuteron wave function, i.e. independent of $c$ (c.f. Eq.~(\ref{13})).
Therefore the cancellation of the real parts of the amplitudes
$f^{(1)VGV}_{\pi d}$ and $f^{(2)}_{\pi d}$ 
can not be achieved in this limit (the value 
$c=2<\!|1/r|\!>_d\approx 1.2\mu$ is obviousely not realistic).

In Ref. \cite{9} we have calculated the sum of the real parts of the diagrams presented in Fig. \ref{2}
varying the parameter $c$ in the limits  $2.5\,\mu\le c\le 3.5\,\mu$.
The results of the present numerical calculation 
are presented in Table 1 for the case when $c$  varies in a larger range 
and the terms $\sim O(\mu/m)$ are taken into account. 
In the calculation we use the purely hadronic values for 
$b_0$ and $b_1$ presented in Ref. \cite{7}, i.e. $b_0=-2.2\times 10^{-3} \ m_\pi^{-1};
b_1=90.5\times 10^{-3} \ m_\pi^{-1}$.
This Table clearly confirmes the conclusion discussed above that the real parts of the
diagrams of Fig. \ref{fig2}b and \ref{fig2}c do not cancel each other.


\section{Summary}

We developed a consequent potential approach to the problem of
the calculation of the pion-deuteron scattering length. The $\pi d$ amplitude
was calculated including terms of the second order with respect to 
the pion-nucleon potential $V_{\pi N}$. 
The proper symmetrization of the wave function for the intermediate nucleons 
is taken into account automatically in our approach.

We show that there is a significant cancellation of the contributions from the imaginary parts of
LCIA (cf. Fig. \ref{2}b) and RE (cf. Fig. \ref{2}c). 
This cancellation is expected. It simply reflects the fact that 
the process $\pi^- d\to \pi^0 nn$  is strongly suppressed near
threshold as required by the Pauli principle.
However,  no such cancellations take place between the real parts of these processes.
The integrals for the real parts of the amplitudes~(\ref{7}) and (\ref{11}) are quite different.
In particular, they have different dependence on the cut off parameter $c$ in the formfactor.
Therefore we see no reasons for the cancellation of
Re$\,f^{(1)VGV}_{\pi^- d}$ and Re$\,f^{(2)}_{\pi^- d}$.

The situation for the reaction \gamde\ is quite analogous to that discussed for 
the reaction $\pi d\to\pi d$. There is no any reasons for the cancellation of the real
parts of the diagrams shown in Fig. \ref{1}a and \ref{1}b.
This conclusion is in agreement with the results of papers~\cite{2,3,4,Beane} where 
rescattering effects are found to be important for the reaction \gamde\ .


We would like to thank A. M. Gasparyan  and V. G. Ksenzov for useful discussions.
This work was partly supported by RFBR grant N$^0$~02-02-16465, DFG-RFBR grant 436 RUS 113/652/0-1.











%




%


\newpage



\begin{table}             
\begin{center}
\vspace*{1.2cm}
\begin{tabular}{|c|c|c|}
\hline
& $Re f_{\pi d}^{(1)VGV}$ [fm] & $Re f_{\pi d}^{(2)}$ [fm]\\
\hline
$c=2\mu$ & 0.0280& -0.0084\\
\hline
$c=3\mu$ & 0.0443  & -0.0098\\
\hline
$c=4\mu$ & 0.0608  & -0.0104\\
\hline
$c=5\mu$ & 0.0774  & -0.0107\\
\hline
\end{tabular}
\label{pirhocontr}
\end{center}
\caption{The real parts of the contributions from the diagrams shown in Fig.\ref{fig2}b and \ref{fig2}c
for the charge exchange process.}
\end{table}

\begin{figure}
\begin{center}
\vspace*{-0.5cm}
\epsfig{file=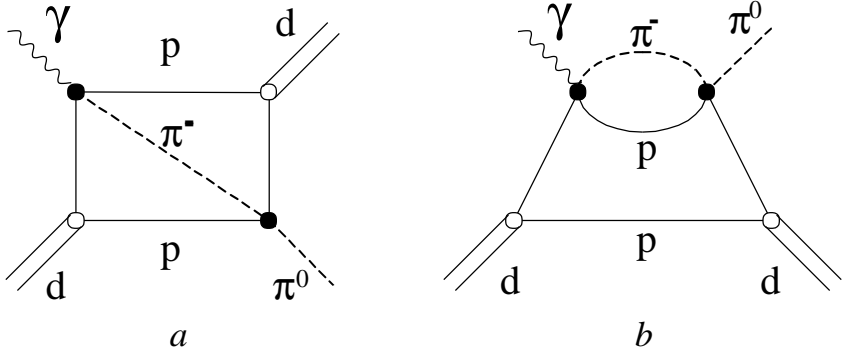,width=11.2cm}
\caption{Diagrams with intermediate negative pion rescattering contributing to the process
$\gamma d\to \pi^0 d$.}
\label{fig1}
\end{center}
\end{figure}

\begin{figure}
\begin{center}
\vspace*{-0.5cm}
\epsfig{file=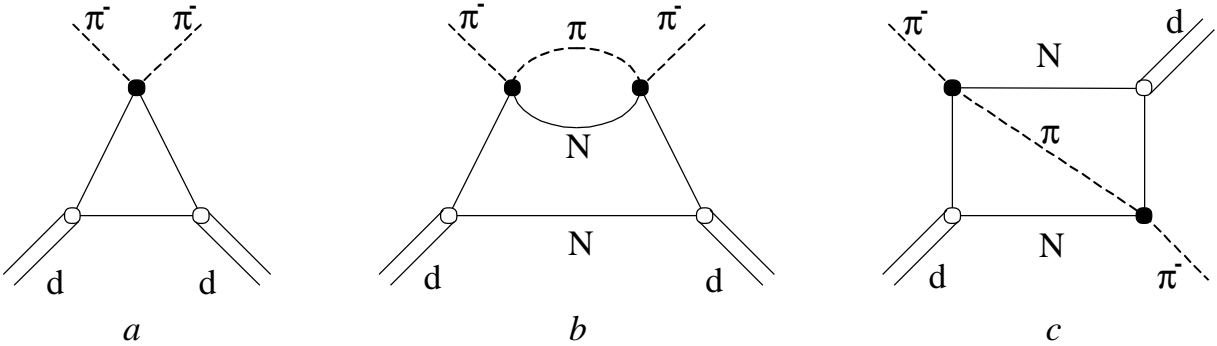,width=16.2cm}
\caption{Feynman diagrams contributing to the $\pi d$-scattering amplitude:
             {\it a} -- diagram of the first order on the $\pi N$ potential;
              {\it b,c} -- diagrams of the second order on the $\pi N$ potential.}
              
\label{fig2}
\end{center}
\end{figure}

\end{document}